\documentclass[%
preprint,
 amsmath,amssymb,
 aps,
]{revtex4-2}

\usepackage{graphicx}
\usepackage{dcolumn}
\usepackage{bm}

\usepackage{appendix}
\usepackage[utf8]{inputenc}
\usepackage{epstopdf, epsfig}
\usepackage{subcaption}
\usepackage{float}
\usepackage{hyperref}
\hypersetup{hypertexnames=false} 
\hypersetup{
    colorlinks=true,
    linkcolor=blue,
    citecolor=blue,
    filecolor=blue,      
    urlcolor=blue,
    linktocpage=True
    }
\usepackage[font=footnotesize]{caption}
\usepackage{siunitx}

\DeclareMathOperator{\erf}{erf}  

\usepackage[textsize=footnotesize, textwidth=2cm, tickmarkheight=5, backgroundcolor=yellow, linecolor=black]{todonotes}

\usepackage{tabularray}
\NewDocumentEnvironment{fancytblr}{+b}{
\begin{tblr}{
  row{1-Z} = {font=\footnotesize},
  hline{1,Z} = {0.08em},  
  hline{3} = {0.05em},
  cells = {c},
  }
#1
\end{tblr}
}{}

\newcommand{\con}{$CON$}
\newcommand{\KP}{$K2P$}
\newcommand{\KPf}{$K2P - \mathbf{u \cdot f}$}
\newcommand{\dtKE}{$D_b E_k / Dt$}

\newcommand{\hotspot}{$\mathbf{x_+}$}

\defcitealias{boettger2023}{B23}

\begin{document}

\title{The energetic inception of breaking in surface gravity waves under wind forcing}

\author{Daniel G. Boettger}
\author{Shane R.  Keating}
\author{Michael L. Banner}
\author{Russel P. Morison}
\author{Xavier Barth\'el\'emy}
\affiliation{School of Mathematics and Statistics, University of New South Wales,
Sydney, Australia}

\date{\today}

\begin{abstract}
The breaking of surface gravity waves is a key process contributing to air-sea fluxes and turbulent ocean mixing. The highly nonlinear nature of wave breaking, combined with the challenges of observing this process in a laboratory or field setting, leaves our understanding of the energetic processes underpinning wave breaking incomplete. Progress towards refining this understanding was made in a recent study (D. G. Boettger et. al., An energetic signature for breaking inception in surface gravity waves, \textit{Journal of Fluid Mechanics} 959, A33 (2023)), which identified an energetic signature in the wave kinetic energy evolution that preceded breaking onset and correlated with the strength of the breaking event. In this study, we examine the influence of wind forcing on this energetic signature. We develop a numerical wave tank that simulates wind flowing over mechanically generated waves and construct an ensemble of cases with varying wave steepness and wind forcing speed. The wind is shown to modulate the wave geometry and elevate kinetic energy at crest tip by up to 35 \%. Despite these influences on the wave, the energetic inception signature was found to robustly indicate breaking inception in all cases examined. At breaking inception, a kinetic energy growth rate threshold was found to separate breaking and non-breaking waves. Under wind forcing, the timing of the energetic inception point occurred slightly earlier than unforced breaking waves, giving advance warning of breaking 0.3 wave periods prior to breaking onset. 
\end{abstract}

\maketitle

\section{Introduction}

The breaking of surface waves is one of the principal mechanisms for the transfer of momentum, energy, heat and gas between the atmosphere and ocean \cite{dasaro2014}. While the implications of these fluxes are typically characterised over large space and time scales, the physical processes leading to wave breaking are highly localised. As a consequence of this contrast of scales, wave breaking cannot be explicitly represented in Earth system models and must be parameterised, however, the accuracy of these parameterisations  \cite{stopa2016, xu2023} is limited by our incomplete understanding of the dynamics underpinning wave breaking \cite{sullivan2010}. 

The breaking process can be conceptually described through the terms breaking inception and breaking onset, where the former describes the initiation of an irreversible process within the crest that leads to breaking, and the latter is the instant when the first surface manifestation of breaking occurs at the crest \cite{derakhti2020}. Inception is a useful concept as it suggests the possibility of identifying a breaking wave in the earlier stages of its evolution when the process is more readily parameterised. 

Efforts to characterise breaking inception have utilised diagnostic parameters based on the kinematic, geometric, or dynamic properties of the wave \cite{perlin2013}. One such approach is based on the diagnostic parameter $B$ \cite{barthelemy2018}, which represents the normalised ratio of the energy flux to the energy density. At the interface, this reduces to the kinematic ratio of particle velocity to crest speed $\| \mathbf{u} \| / \| \mathbf{c} \|$. A range of laboratory and numerical studies \cite{saket2017,saket2018,derakhti2018,seiffert2018, derakhti2020,touboul2021} have reported that a threshold value $B_{th} = 0.855 \pm 0.05$ exists that separates breaking and non-breaking waves such that if $B_{th}$ is exceeded the crest will always evolve to break. Furthermore, it has been shown \cite{derakhti2020,na2020} that the parameter
\begin{equation}\label{eq:Gamma}
    \Gamma = T \left. \frac{D_b B}{Dt}\right|_{B_\mathrm{th}},
\end{equation}
formulated as the rate of change of  $B$ as it passes through $B_{th}$ normalised by the wave period $T$, accurately predicts the breaking strength parameter $b$ \cite{phillips1985}, which has been shown to quantify the energy dissipated through breaking \cite[e.g.][]{drazen2008, Deike2015, sutherland2015}. 

While the approach of \cite{barthelemy2018} has dynamical foundations and has proven to be a robust indicator of breaking inception, it does not shed light on the underlying dynamical processes that cause the wave to break. This motivated \citet{boettger2023} (hereafter \citetalias{boettger2023}) to identify a new breaking inception threshold associated with the energetics, rather than the kinematics, of breaking waves. \citetalias{boettger2023} investigated the evolution of the energetic properties of breaking waves in the context of a kinetic energy balance equation,
\begin{equation}\label{eq:DbEkDt-balance}
\frac{D_b E_k}{D t}= \underbrace{- \nabla \cdot \left(\mathbf{u} p + [\mathbf{u-b_+}]E_k \right)}_{\text{\con{}}} 
    \underbrace{- \rho g w}_{\text{\KP{}}} 
    + \underbrace{\mathbf{u \cdot f}}_{\text{friction}} + \underbrace{\mathbf{u \cdot n} \sigma \kappa \delta_s}_{\text{sfc tension}},
\end{equation}
which tracks the rate of change of the kinetic energy density $E_k$ at its local maxima  $\mathbf{x_+}$ near the crest tip. The operator $D_b E_k /Dt = \partial E_k/ \partial t + \mathbf{b_+} \cdot \nabla E_k$ captures the local and convective components of the $E_k$ evolution \cite[][eq. 2.2]{tulin-2007}, with $\mathbf{b_+}=d \mathbf{x_+} / d t$. \citetalias{boettger2023} found that (\ref{eq:DbEkDt-balance}) is dominated by the convergence of kinetic energy \con{} and its conversion to potential energy \KP{}, while surface tension is negligible and friction is only significant at breaking onset. In non-breaking waves, these source and sink terms are approximately balanced, so that $D_b E_k /Dt$ is weakly positive during wave growth and negative during wave decay. For waves that go on to break, however; this balance is disrupted by a rapid increase in \con{} that is not offset by a corresponding increase in the magnitude of \KP{}, leading to rapid growth in $E_k$. This energetic signature was shown to occur up to 0.4 wave periods prior to breaking onset, with the magnitude of $D_b E_k /Dt$ at this instant also correlated with breaking strength through the parameter $\Gamma$. Hence, this energetic inception threshold both predicts the occurrence of breaking onset and indicates the strength of the breaking event. 

In the open ocean, the growth of the wave field is driven by wind forcing, which injects kinetic energy through both form drag and viscous stress components \cite{melville1996, sullivan2010}. As well as being a primary source of wave growth, wind forcing modulates the characteristics of breaking waves. The geometric modulation is well reported: the height of the wave is increased (decreased) under following (opposing) winds, while the horizontal and vertical asymmetry of the wave increases with wind speed \cite{leykin1995, Reul2008, chen2022}. The energetic modulation is less well understood, which is partly due to the challenges of measuring the wave energetic properties in experimental settings. Several authors have reported that the time of breaking onset is accelerated in the presence of wind, with the majority of this impact attributed to the wind drift current as opposed to the wind-induced pressure modulations \cite{banner2002, chen2022}. The effect also appears to be dependent on the wave age  $c_p / u_*$ \cite{waseda1999, Iafrati2019}, with energetic growth rates suppressed in old waves but enhanced for younger waves. These results allude to the complexity of the energetic processes underlying wave breaking and which are yet to be fully explored. 

The study of \citetalias{boettger2023} provides unique insights into the dynamical evolution of a wave crest through breaking inception and breaking onset, but it is unclear whether the energetic inception threshold remains robust for the more realistic case of wind forced waves that are characteristic of the open ocean. In this study, we investigate how the terms in the $E_k$ balance equation (\ref{eq:DbEkDt-balance}) evolve in wind forced waves. We do so by developing a two-phase numerical wave tank that accurately simulates the interaction between the wind and the wave field, while being computationally efficient enough to enable us to sample a wide range of wave and wind forcing magnitudes. The resultant ensemble of wind-forced waves enables us to present a comprehensive understanding of the role of wind forcing on energetic breaking inception. 

\section{Experimental details}\label{s:exp-details}

\begin{figure}
  \includegraphics[width=\textwidth]{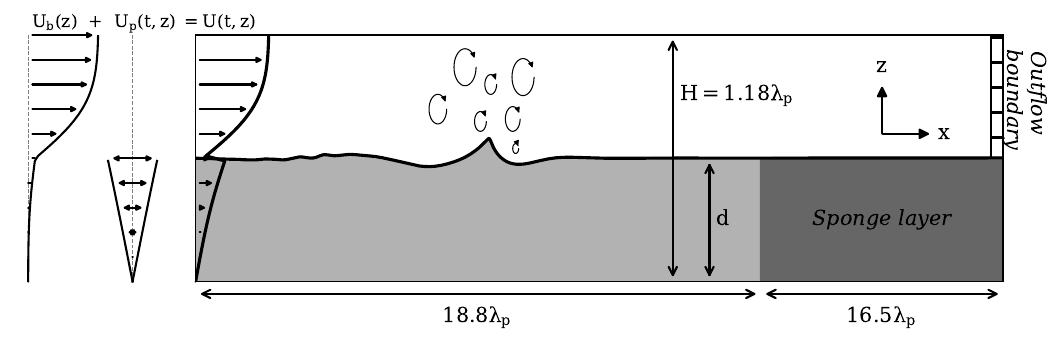}
  \caption{Schematic of our numerical wave tank. The boundary forcing $\mathbf{U}(t,z)$ is a combination of a time-independent boundary layer velocity $\mathbf{U_b}(z)$ and the time-varying paddle velocity $\mathbf{U_p}(t,z)$. Waves are generated at the left hand boundary and travel down the tank in the positive $x$ direction before being absorbed by the numerical sponge layer. A typical chirped wave (enlarged for clarity) is shown, with tank dimensions normalised by the deep-water wavelength $\lambda_p$ derived from the paddle frequency $\omega_p$.}
\label{fig:tank}
\end{figure}

We use the Navier-Stokes solver Gerris \cite[][]{popinet2003, popinet2009} to simulate a two-dimensional Numerical Wave Tank (NWT) and generate an ensemble of non-breaking and breaking waves under the influence of a range of wind speeds. Gerris implements the volume-of-fluid method to simulate two-phase, incompressible flow, including the effects of surface tension and viscosity. Adaptive grid refinement enables highly complex flows to be efficiently simulated. Gerris has been used to examine numerous aspects of gravity waves, including waves breaking on a beach \cite{Wroniszewski2014}, wave breaking kinematics \cite{deike2017,pizzo2016}, energy dissipation \citep{devita2018} and wind effects \cite{Iafrati2019}. The NWT setup (Fig. \ref{fig:tank}) mimics laboratory experiments in which a fan drives a steady wind flow and waves are mechanically generated by a paddle \cite[e.g.][]{banner1990, reul1999, peirson2008, grare2013}. The configuration is based on that utilised by \citetalias{boettger2023}, extended to enable the application of a realistic air-water boundary layer profile. Here, we provide an overview of the features of the NWT and the method utilised to implement the wind forcing and refer the reader to \citetalias{boettger2023} for further details. 

The domain is configured in non-dimensional coordinates scaled by the wavelength $\lambda_p$ and period $T_p$ of the central wave in the wave packet, such that the tank has a height $H$ of $1.18\lambda_p$ and a useable length of $18.8\lambda_p$ (Fig. \ref{fig:tank}). The depth of water is set to $d/\lambda_p = 0.59$ so that there is negligible interaction with the bottom of the NWT. The numerical domain extends a further $16.5\lambda_p$ to accommodate a numerical sponge layer, which effectively absorbs the wave packet energy by both gradually increasing viscosity and decreasing model grid resolution. The incoming wind flow is matched by an outflow boundary condition for the air phase, enabling the wind to exit the simulation domain. 

To reduce computational cost, the domain is configured as a two-dimensional ($\mathbf{x} = (x,z)$) NWT. Previous studies \cite{devita2018,Derakhti2016,barthelemy2018} have reported no significant difference in the integrated wave energetics between two- and three-dimensional simulations. Furthermore, the characteristics of the wind-wave momentum and energy flux can also be sufficiently represented in two dimensions \cite{Reul2008}. Limiting the study to two-dimensional simulations therefore allows a wide range of wave parameters to be examined over a large ensemble within computational constraints, while still accurately capturing the physical characteristics of the waves and airflow above.

Waves are generated at the left boundary of the NWT by applying the velocity and pressure gradient solutions for a bottom-mounted, flexible flap paddle from wavemaker theory \cite{dean1991}. A chirped packet function \cite{song2002} is utilised to generate a compact, focusing wave packet with number of waves in the paddle signal $N=5$, chirp rate $C_{ch}=1.0112 \times 10^{-2}$ and a paddle frequency $\omega_p$ equivalent to the orbital frequency of a deep water gravity wave with wavelength $\lambda_p=1$ \si{\metre} and linear phase speed $c_p=1.15$ \si{\metre\per\second}. 

Adaptive grid refinement is used to focus fine resolution in regions of high vorticity and at the air-water interface. Each level of refinement divides the parent cell into four, resulting in a maximum resolution equivalent to a uniform mesh with $2^n \times 2^n$ grid cells, for $n$ refinement levels. As our focus is on the near-surface wave energetics, we assess the maximum refinement level required using the viscous boundary layer thickness of the water phase $\delta \approx{\lambda_p}/{\sqrt{\text{Re}}}$ \cite[eq. (5.7.4)]{batchelor_2000}, where $\text{Re}=\rho_w c_p\lambda_p / \mu_w$ is the wave Reynolds number defined using the density $\rho_w$ and dynamic viscosity $\mu_w$ of the water phase. We set $\text{Re} = 4 \times 10^4$, which enables the boundary layer to be resolved by  approximately four cells at a refinement level of $2^{10}$ (equivalent to a grid resolution $dx = \lambda_p/870$) and was shown by \citetalias{boettger2023} to achieve numerical convergence. While the chosen $\text{Re}$ is smaller than the physical $\text{Re} \approx 1 \times 10^6$ for a deep water gravity wave, it has been shown \cite{deike2017, mostert2020} to be large enough that viscous effects are not dominant and all energy within the boundary layer is adequately resolved. 

Particular attention is given to the simulation of the air-water boundary layer, the shape of which is critical in determining the evolution of the wind-driven wave field. There are two main aspects of the boundary layer that characterise these processes. The first of these is the wave age $c_p / u_*$, which is the ratio between the wave phase speed and the airside friction velocity $u_* = \sqrt{\vert {\tau} \vert / \rho_a}$ , where $\vert {\tau} \vert$ is the magnitude of the viscous shear stress at the air-water interface and $\rho_a$ the air density. The wave age signifies the balance of forces between the wind and the wave field. The wind and wave field are in equilibrium when  $c_p / u_* \sim 30$, with values smaller (greater) than this typical of growing (decaying) seas \cite{sullivan2010}. 

The second feature of importance is the wind drift layer in the water. This takes the form of an exponentially decaying profile with a surface velocity $U_s \sim 3 \%$ of the free stream air velocity $U_a$ \cite{banner1998, liberzon2011, longo2012, buckley2020}. The shape of the wind drift layer has a significant impact on the non-linear wave-wave interactions \cite{nwogu2009}. In a numerical study using a two-phase solver similar to that implemented in our NWT, \citet{Zou2017} found that an exponential wind drift layer was necessary to accurately replicate wave evolution, while an equivalent linear shear or uniform current layer misrepresented the wave amplitude and breaking location. 

In a two-phase numerical simulation of wind-forced waves the proper representation of the boundary layer is typically achieved by initialising the simulation from rest and allowing the boundary layer to develop from an initially uniform wind flow \cite{Iafrati2019, wu2022}. This, however, requires a large domain with sufficient fetch or the use of periodic boundary conditions. In both cases, the duration of the initialisation period may extend over $O(100)T_p$, which makes the simulation of a large ensemble of breaking waves computationally challenging. 

To overcome these challenges, we utilise a two-stage modelling approach similar to \citet{sullivan2018a} in which the air-water boundary layer is computed in a precursor simulation and then applied as an initial and boundary condition for subsequent experiments. In this first stage, the NWT is run without any paddle forcing and the air-water boundary layer evolves from an initially quiescent state following the application of a steady, uniform air flow. An analytical solution of the form \cite{lock1951, boeck2005}
\begin{equation}\label{eq:boundary-layer-erf}
    U_b(z) = 
    \begin{cases}
        (U_a - U_s) \erf \left( \frac{z-\eta}{\delta_a} \right) + U_s & (z \geq \eta) \\
        (U_s - U_w) \erf \left( \frac{z-\eta}{\delta_w} \right) + U_s & (z < \eta)
    \end{cases}
\end{equation}
is then fit to the velocity profile. Here, $\delta_a$, $\delta_w$ are the thickness of the air and water boundary layers, $U_a$, $U_w$ are the undisturbed horizontal air, water velocities and $U_s$ is the tangential velocity at the air-water interface $z=\eta$. For an initial uniform air velocity of  $U_0 / c_p = [1,2,6]$, the resultant boundary layer profile has a surface velocity $U_s$ between 2.3 \% and 3.1 \% of the free stream air velocity $U_a$ and a wave age range spanning moderate ($c_p / u_* \approx 21.1$) to strong ($c_p / u_* \approx 8.2$) wind forcing (table \ref{tab:experiments}). The turbulent characteristics of the air boundary layer for the simulated wind forcing speeds can be quantified using the air Reynolds number $\text{Re}_a = (\rho_a U_0 [H-d])/\mu_a$ and the friction Reynolds number $\text{Re}_* = (\rho_a u_* [H-d])/\mu_a$, which utilise the height of the air phase $[H-d]$ as the length scale and either $U_0$ or $u_*$ as the velocity scale. The magnitude of these parameters (table \ref{tab:experiments}) spans fully laminar to weakly turbulent flow \cite{Kim1987, wu2022}.
\begin{table}[]
    \centering
    \begin{fancytblr}
    \SetCell[c=7]{} Boundary layer & & & & & & & & Paddle & \SetCell[c=3]{} Simulations & & \\
    $U_0/c_p$ & $c_p / u_*$ & $\text{Re}_a$ & $\text{Re}_*$ & $U_s / c_p$ & $U_a / c_p$  & $\delta_a / \lambda_p$ & $\delta_w / \lambda_p$ & $A_p/\lambda_p$ & $n_\text{sim}$ & $n_\text{b}$ & $n_\text{nb}$ \\
    0 & $\infty$ & 0 & 0 &  0 & 0 & 0 & 0                 & $0.025-0.050$ & 63  & 28 & 289 \\ 
    1.0 & 21.1 & 1200 & 60 & 0.034 & 1.50 & 0.361 & 0.132 & $0.040-0.044$  & 26  & 14 & 144 \\
    2.0 & 17.3 & 2500 & 80 & 0.078 & 2.99 & 0.356 & 0.172 & $0.040 - 0.044$ & 39  & 21 & 152 \\
    6.0 & 8.2 & 7400 & 150 & 0.259 & 8.31 & 0.295 & 0.169 & $0.043 - 0.050$ & 45  & 36 & 144 \\
    \hline
    &   &    & & & & & & \SetCell[r=1]{r} Total: & 173 & 99 & 729 \\    
    \end{fancytblr}
    \caption{Summary of experiments included in this study. The NWT was configured using a range of boundary layer profiles derived from a uniform wind forcing $U_0 / c_p$ and described analytically with (\ref{eq:boundary-layer-erf}). The characteristics of the wind profile are defined with the wave age $c_p / u_*$, air Reynolds number $\text{Re}_a$ and friction Reynolds number $\text{Re}_*$. For each configuration the amplitude of the paddle $A_p/\lambda_p$ was varied over $n_\text{sim}$ simulations to generate an ensemble of $n_\text{b}$ breaking and $n_\text{nb}$ non-breaking crests. All values are normalised by the linear wavelength $\lambda_p$ and phase speed $c_p$ of the paddle frequency $\omega_p$. Simulations with $U_0 / c_p=0$ are also utilised in \citetalias{boettger2023}.}
  \label{tab:experiments} 
\end{table}

Subsequent simulations are then initialised using (\ref{eq:boundary-layer-erf}) and the appropriate parameters from table  \ref{tab:experiments}. The left-hand velocity and pressure boundary conditions of the NWT become the sum of (\ref{eq:boundary-layer-erf}) and the chirped packet function described in \citetalias{boettger2023}. The paddle remains stationary for the first $10 T_p$ of each simulation to allow the turbulent characteristics of the boundary layer to develop, after which it continues for a further $15 T_p$ - $20 T_p$ as the wave packet propagates through the NWT. This approach enables a realistic air-water boundary layer with balanced interfacial shear stress to be simulated in a computationally efficient manner and negates the requirement for long initialisation periods in every experiment.

For each wind forcing speed, a series of simulations are conducted in which the amplitude of the paddle signal is set within the range $A_p / \lambda_p \in [0.025, 0.050]$ to generate a combination of non-breaking and breaking waves of varying steepness and breaking strength. Given the finite resolution of the simulations, we define a crest as breaking if the interface contour exceeds the vertical by a horizontal distance $d\eta_x \ge 0.5 {dx}$ over a length $d\eta_z \ge {dx}$, and the instant of breaking onset as the first time that these thresholds are exceeded. Over the wind speed range $U_0 / c_p \in [0,6]$ the total ensemble consists of 99 breaking and 729 non-breaking waves (table \ref{tab:experiments}). The $U_0 / c_p =0$ simulations in this ensemble are the same as the $N=5$, $d/\lambda_p = 0.59$ cases from \citetalias{boettger2023}, enabling a point of comparison between these two studies. 

With the objective of determining how the local crest energetic properties are modulated in the presence of wind forcing, our analysis is focused on an examination of the evolution of the energetic terms in (\ref{eq:DbEkDt-balance}). We characterise the local crest energetics at the location $\mathbf{x_+} = [x_+(t), z_+(t)]$ where the local kinetic energy density $E_k$ has its maximum value and which moves with velocity $\mathbf{b_+}=d \mathbf{x_+} / d t$. To enable comparison of the crest energetics across all waves in the ensemble, a crest reference location and time are set as $[x_0, t_0] = [x_+, t]$ at the instant of breaking onset for breaking crests and at the instant of maximum $E_k$ for non-breaking crests. The evolution of the crest in space and time is then referenced to these parameters using the non-dimensional coordinates $x^* = (x - x_0) / \lambda_p$,  $z^* = z / \lambda_p$ and $t^* = (t - t_0) / T_p$ utilising the wavelength $\lambda_p$ and period $T_p$ derived from the paddle forcing frequency $\omega_p$, which is constant for all simulations. 

\section{Airflow and wave characteristics}

\begin{figure}
    \centering
    \includegraphics[trim={0 0.5cm 0 0},clip,width=\textwidth]{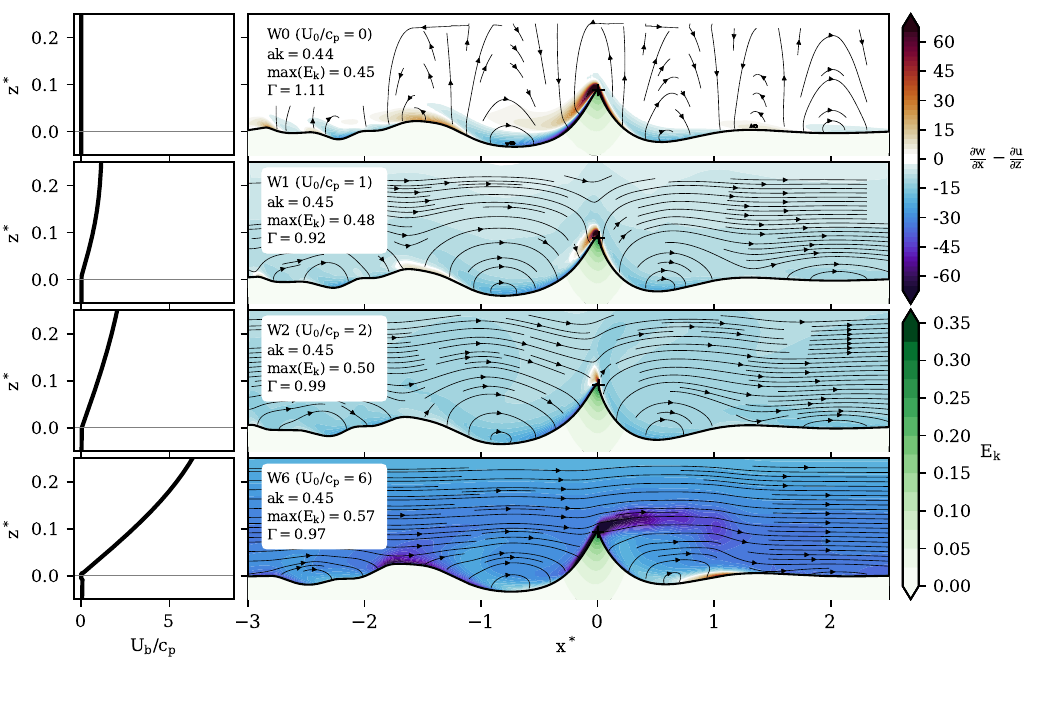}
    \caption{(left) The boundary layer profile applied at the left-hand boundary of the simulation domain for (top to bottom) $U / c_p = [0,1,2,6]$. (right) The air vorticity $\partial w / \partial x - \partial u / \partial z$ (referenced to upper colorbar) and water kinetic energy $E_k$ (lower colorbar) field for the central wave in the (top to bottom) W0, W1, W2 and W6 packets at breaking onset ($t^*=0$). Streamlines indicate the absolute air velocity $\mathbf{u_a}$. An animated version of this figure is provided as supplementary material.}
    \label{fig:wind-snapshot-combined}
\end{figure}

To illustrate the unique characteristics of the air flow and the wave field at each wind speed, we choose four breaking waves with similar maximum wave steepness $\max(ak) \approx 0.45$. These cases are shown in Fig. \ref{fig:wind-snapshot-combined} at the instant of breaking onset ($t^* =0$) and labelled (top to bottom) W0, W1, W2 and W6 respectively (an animation of this figure is provided as supplementary material). The air flow is characterised by the vorticity $\partial w/\partial x - \partial u / \partial z$ and velocity streamlines. In the absence of wind forcing (W0, Fig. \ref{fig:wind-snapshot-combined} top), the airflow is dominated by wave-induced motions characterised by regions of positive (negative) vorticity at the wave crest (trough). As the wind speed increases, the wave-induced modulations are exceeded by the wind forcing and replaced by a vertical shear layer characterised by negative vorticity of increasing magnitude with wind strength. At the strongest wind forcing $U_0/c_p=6$ airflow separation is evidenced by the stream of enhanced negative vorticity emanating from the crest tip and the reversal of flow in the downstream wave trough. 

While the wave steepness $ak$ in each case is similar, the symmetry of the wave clearly changes. As the wind speed increases, the wave crest can be seen to lean further forwards such that the forward face steepness exceeds that of the rear face, consistent with previous studies \cite{leykin1995, Reul2008, chen2022}. Wind forcing also increases the wave kinetic energy density, such that $E_k$ at \hotspot (denoted by the $+$) for the W6 case is 26 \% larger than the equivalent W0 wave. Conversely, the breaking strength parameter $\Gamma$ does not show a clear correlation with wind speed or $E_k$, with the W0 wave being a more strongly breaking case then its W6 counterpart. While the addition of wind forcing clearly increases the absolute energy of the waves, the underlying processes leading to wave breaking are characterised by the rate of wave growth (e.g. (\ref{eq:Gamma})) and not necessarily the absolute values. This motivates our analysis of the rate of wind energy input to the waves in the following section.  

\section{Wind energy input}

The rate of energy input from the wind to the water
\begin{equation}\label{eq:S-in}
    S_\text{in} = S_f + S_\nu = -p_s \mathbf{n} \cdot \mathbf{u_s} + \boldsymbol{\tau_\nu} \cdot \mathbf{u_s}
\end{equation}
consists of two components. The form drag contribution $S_f$, arising from the pressure differential caused by the flow of the wind over the wavy surface, is described by the alignment between the surface pressure $p_s$ and the normal component of the surface velocity $\mathbf{u_s}$. The viscous stress component $S_\nu$ is a function of the shear stress vector $\boldsymbol{\tau_\nu} = \mu_a \left(\nabla \mathbf{u} + \nabla \mathbf{u}^T \right) \cdot \mathbf{n}$ and $\mathbf{u_s}$. It can be further decomposed into the component of $\boldsymbol{\tau_\nu}$ aligned with the wave orbital velocity, which contributes to wave growth, and the component aligned with the drift velocity, which enhances the wind drift layer. For steep waves such as those in our ensemble, the majority of $S_{\nu}$ goes towards the wind drift layer and $S_f$ is primarily responsible for wave growth \cite{peirson2008, buckley2020, wu2022}. While our experiments each run over a relatively short time period, we are able to accurately simulate the distribution of energy into both the wind drift layer and the wave growth through the application of the boundary condition $U_b$ (\ref{eq:boundary-layer-erf}) outlined in section \ref{s:exp-details}.

\begin{figure}
 \centering
    \includegraphics[width=1\linewidth]{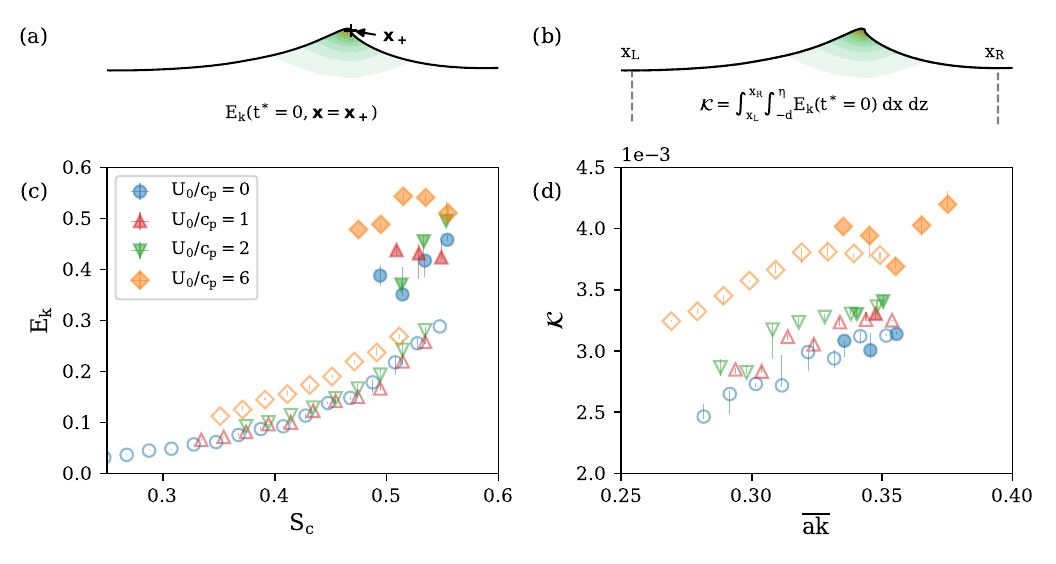}
    \caption{The wave kinetic energy $\mathcal{K}$ is integrated over the area between the troughs $x_L$, $x_R$ and over the full water depth (a). The kinetic energy $E_k$ is measured at its spatial maxima \hotspot{} (b). Each parameter is shown for time $t^*=0$ as a function of wave steepness $ak$ (c) or crest steepness $S_c$ (d). Data are binned by breaking (solid symbols) or non-breaking (hollow symbols) and over intervals of 0.02 to represent the median (symbol) and interquartile range (whiskers) over each bin.}
    \label{fig:wave-energy-input}
\end{figure}

The resultant kinetic energy field for the wind-forced waves therefore contains contributions from mechanical generation, the drift current and the instantaneous wind input. To quantify the magnitude of the wind energy input relative to the hydrodynamic input from the chirped packet, we consider both the local kinetic energy density $E_k$ at \hotspot{} (Fig. \ref{fig:wave-energy-input}(a)) as well as the total wave kinetic energy $\mathcal{K}$ (Fig. \ref{fig:wave-energy-input}(b)) for each wave in our ensemble. The total wave energy is calculated as
\begin{equation}
   \mathcal{K} = \int _{x_L}^{x_R}\int_{-d}^\eta E_k(t^*=0) \; dx \; dz \;,
\end{equation}
where the integral limits are set to the wave trough locations $[x_L, x_R]$ and extend vertically from the bottom of the NWT $z=-d$ to the interface $z=\eta$. While steepness is often used to characterise waves, the appropriate formulation is dependent on the situation. We find the strongest correlation between $\mathcal{K}$ and the time-averaged wave steepness $\overline{ak}$, while the local $E_k$ is best exemplified using the crest steepness $S_c = a \pi / \lambda_c$, where  $\lambda_c$ is the distance between wave zero-crossings \cite{barthelemy2018, banner2014}. To increase the clarity of Fig. \ref{fig:wave-energy-input}(c,d) given the large number of individual waves within our ensemble, we bin this data at 0.02 intervals and display the median and interquartile range for each bin. 

The local kinetic energy is significantly enhanced in breaking waves, with $E_k$ up to twice the magnitude compared to a non-breaking wave of similar steepness. These elevated levels of kinetic energy are, however, highly localised at the crest tip, with \citetalias{boettger2023} finding that the difference between the integrated kinetic energy of breaking and non-breaking waves diminishes if the integral domain exceeds the upper 20 \% of the crest. Accordingly, there is little difference between $\mathcal{K}$ for breaking and non-breaking waves of similar steepness in Fig.  \ref{fig:wave-energy-input}, which contrasts with $E_k$ and demonstrates the utility of the local energetic quantities in understanding breaking wave energetics. Conversely, the impact of the wind forcing on the wave kinetic energy is clearly evident in both the local and integrated magnitude of kinetic energy, with $\mathcal{K}$ systematically increasing by up to 20 \% and $E_k$ by up to 35 \% from $U_0 / c_p = 0$  to $U_0 / c_p = 6$ for waves of equivalent steepness.

\section{Energetic breaking inception}

\begin{figure}
    \centering
    \includegraphics[width=\textwidth]{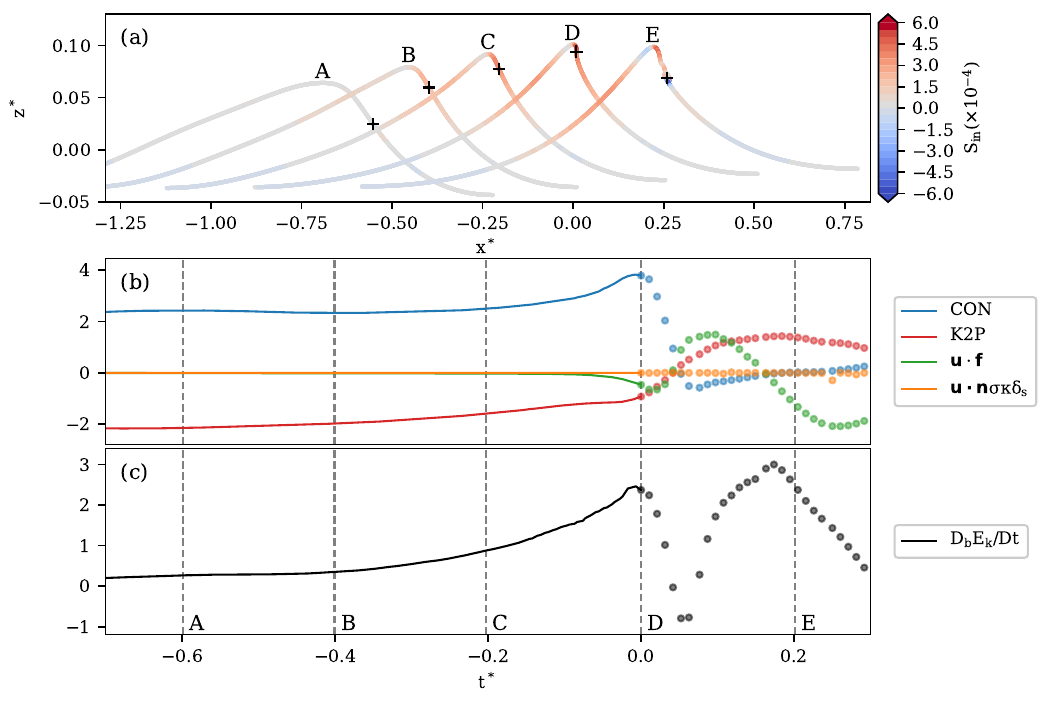}
    \caption{Evolution of the energetic terms for the central wave in the W6 packet. The crest interface, coloured by the local wind energy input $S_\text{in}$ is shown in (a) at intervals of $\Delta t^* =0.2$. The value of terms on the RHS of the $E_k$ balance equation (\ref{eq:DbEkDt-balance}) at the location \hotspot{} (indicated by the $+$) are shown in (b) and their sum $D_b E_k / Dt$ in (c).}
    \label{fig:energy-example}
\end{figure}

We explore the energetic evolution of the representative $U_0 / c_p = 6$ breaking wave, W6, in Fig. \ref{fig:energy-example}. Snapshots of the wave interface and the instantaneous wind energy input $S_\text{in}$ are shown in Fig. \ref{fig:energy-example}(a) at intervals of $\Delta t^* =0.2$. The rate of energy input varies significantly over the wave evolution but is typified by large positive values in the crest tip region and weakly negative values in the troughs that are associated with the reversal of the airflow at these locations (Fig. \ref{fig:wind-snapshot-combined}). The action of the wind on the wave interface is therefore complex and cannot be represented by the local value at \hotspot{}. Instead, we consider  the effect of  $S_\text{in}$ to be included as a component of the kinetic energy flux through the wave (see e.g. Fig. 7(a) in \citetalias{boettger2023}) and we focus our analysis specifically on the energetic terms in  (\ref{eq:DbEkDt-balance}).  These are shown in  Fig. \ref{fig:energy-example}(b), with the resultant growth rate  ${D_b E_k}/{D t}$  in Fig. \ref{fig:energy-example}(c). 

\begin{figure}
    \centering
    \includegraphics[width=0.98\textwidth]{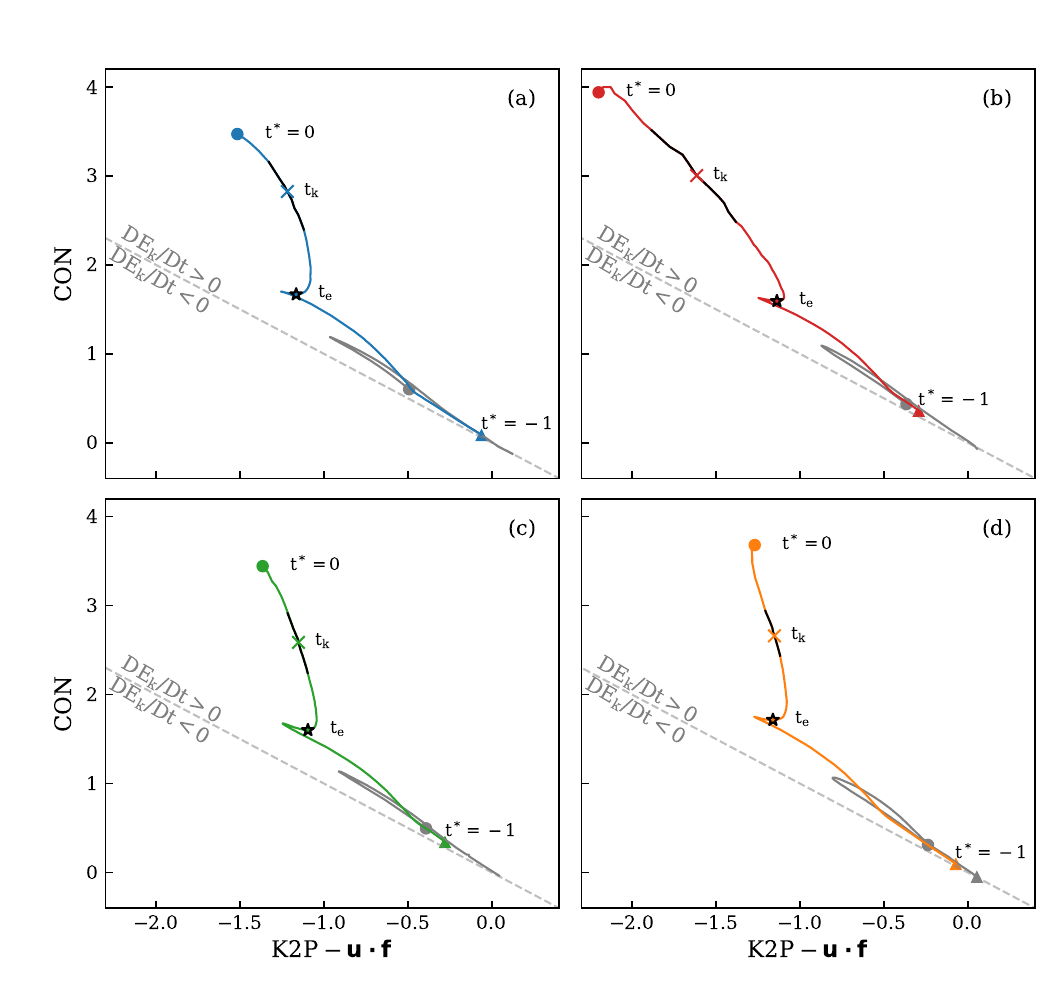}
    \caption{Relationship between the two leading terms in the $E_k$ balance equation (\ref{eq:DbEkDt-balance}) for the W0 (a), W1 (b), W2 (c) and W6 (d) representative breaking crests. In each panel, the preceding non-breaking crests from that packet is also shown (grey line). Values above (below) the dashed line indicate an increase (decrease) in \dtKE{}. The time of the energetic inception threshold ($t_e$) and the kinematic inception threshold ($t_k$) are annotated, with the superposed black lines indicating the period for which $0.83 < B < 0.88$.}
    \label{fig:phase-diagram}
\end{figure}

As reported by \citetalias{boettger2023} for the $U_0 / c_p = 0$ cases, the kinetic energy budget (\ref{eq:DbEkDt-balance}) is dominated by the \con{} and \KP{} terms, with the surface tension term $\mathbf{u \cdot n} \sigma \kappa \delta_s$ negligible and friction $\mathbf{u \cdot f}$ only significant just prior to breaking onset ($t^* = 0$). Prior to $t^*=-0.3$,  ${D_b E_k}/{D t}$  increases slowly as the input of kinetic energy to the crest tip from the \con{} term is largely offset by its conversion to potential energy through the \KP{} term. Beyond this time, the magnitude of \con{} begins to increase rapidly, while the magnitude of \KP{} steadily decreases as the wave approaches its maximum amplitude. As a consequence, ${D_b E_k}/{D t}$ also increases rapidly up to ($t^* = 0$). 
The interplay between the $E_k$ source and sink terms is more clearly seen in Fig. \ref{fig:phase-diagram}, where the relation between \con{} and \KPf{} for the time interval $t^*=[-1,0]$ is shown for the representative waves (W0 to W6) at each wind forcing. In this phase space, the magnitude of the growth rate \dtKE{} is represented by the distance from the dashed line (neglecting the negligible impact of surface tension). In each case, the initial increase in \con{} is offset by a corresponding decrease in \KPf{} such that \dtKE{} remains small. At some time prior to breaking onset, an imbalance is seen to develop between these terms (labelled $t_e$) at which point the increase in \con{} is no longer matched by a corresponding decrease in \KPf{}. This critical point, termed the energetic breaking inception threshold by \citetalias{boettger2023}, is distinct from the kinematic breaking inception threshold \cite{barthelemy2018, derakhti2020}, which, at the wave surface is quantified by the magnitude of the breaking inception parameter $B = \vert \mathbf{u} \vert / \vert \mathbf{c} \vert$ exceeding a threshold value of $B = 0.855 \pm 0.05$. The timing of the kinematic inception threshold $t_k$, shown in Fig. \ref{fig:phase-diagram}, occurs after the energetic inception threshold and is clearly separated in phase space from from $t_e$.

\begin{figure}
    \centering
    \includegraphics[width=0.98\textwidth]{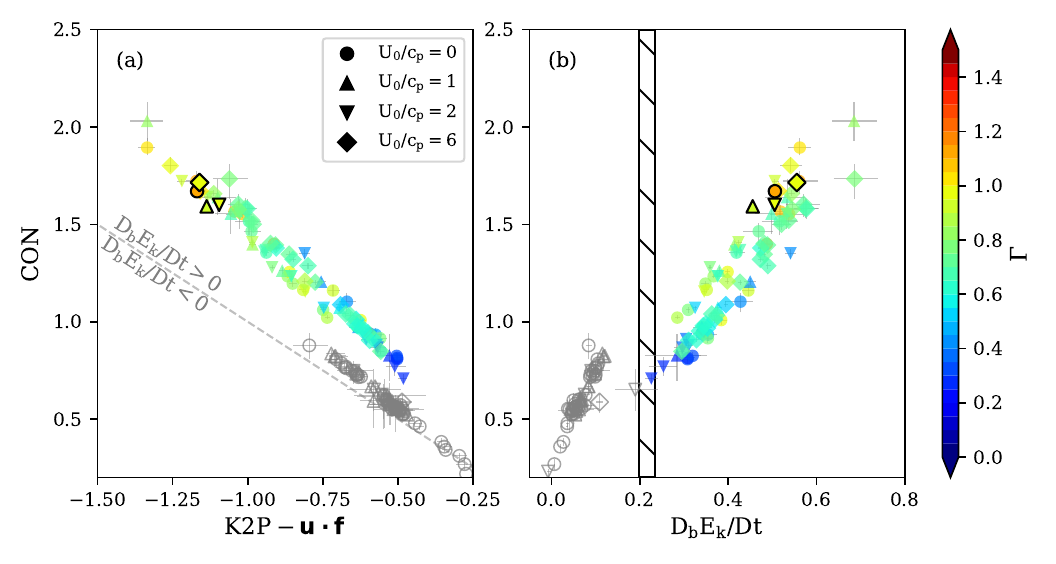}
    \caption{The magnitude of \con as a function of (a) \KPf{} and (b) \dtKE{}, at the time that the critical point occurs. The $5 \%$ and $95 \%$ confidence intervals are also shown. Breaking crests are coloured by the breaking strength indicator $\Gamma$, and non-breaking crests are coloured grey. The \dtKE{} threshold separating non-breaking and breaking crests from \citetalias{boettger2023} is shown by the hatched region. The representative breaking crests W0 to W6 discussed in the text are indicated with black bordered markers.}
    \label{fig:div-vs-rho-summary}
\end{figure}

We define this critical point as the final local minimum of the parametric curve $(x,y) = ([K2P - \mathbf{u \cdot f}], \; CON)$ that occurs before $t^*=0$. We found this to be a generic feature of all breaking waves examined in our ensemble. The location of this point in \con{} and \KPf{} phase space for all of the waves in table \ref{tab:experiments} are shown in Fig. \ref{fig:div-vs-rho-summary}(a), where the error bars represent the 5 \% and 95 \% confidence intervals calculated using a time series bootstrap method. Despite the additional energy input increasing crest $E_k$ magnitudes by up to 35 \%, the waves under the influence of wind forcing fall on the same phase space plane as those without wind forcing. 

A key detail of the energetic inception threshold is that a critical point in the evolution of \con{} is a necessary but not sufficient condition preceding breaking onset. Critical points are identified in some non-breaking cases (grey markers); however, the growth rate \dtKE{} in these cases is small. Consistent with \citetalias{boettger2023}, we find that a threshold region exists of $D_b E_k / Dt = [0.198, 0.235]$ within the confidence intervals of the data (Fig. \ref{fig:div-vs-rho-summary}(b)). Below this threshold region a critical point in \con{} does not lead to breaking, but all waves in our ensemble for which a critical point is observed above this threshold proceed to break without exception. The energetic inception threshold can therefore be considered an energetic tipping point: for $ D_b E_k / Dt \geq 0.235$ any further increase in the rate of kinetic energy convergence at the crest tip will lead to breaking onset. 

For each breaking wave we also estimate the breaking strength with the parameter $\Gamma$ using (\ref{eq:Gamma}), where we calculate ${D_b B}/{Dt}$ as the average rate of change over the interval $B\in[0.83,0.88]$ and $T$ is defined from the deep water wave relationship using the crest zero-crossing wavelength $\lambda_c$. In Fig. \ref{fig:div-vs-rho-summary} we show that $\Gamma$ increases with increasing magnitude of \con{} and $D_b E_k / Dt$ at the time of energetic inception. This result indicates that the strength of the breaking event can be estimated using the energetic quantities measured at the time of energetic breaking inception. 

\begin{figure}
    \centering
    \includegraphics[width=\textwidth]{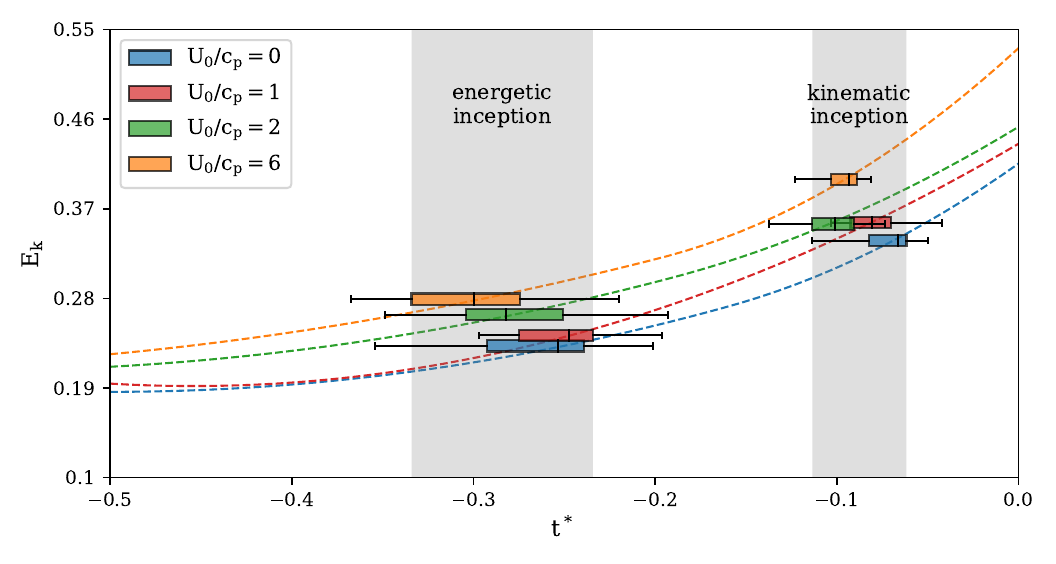}
    \caption{The energetic ($t_e$) and kinematic ($t_k$) inception times for all breaking crests in our ensemble, relative to the time of breaking onset $t^* = 0$. For each wind speed, times are represented by their quartiles (box) and the minimum/maximum values within 1.5 times the interquartile range (whiskers). A representative $E_k$ evolution (dashed lines) is computed by fitting a line to the median $E_k$ values at $t^* = [-1, t_e, t_k, 0]$.}
    \label{fig:inception-times}
\end{figure}

Recall that the energetic inception threshold clearly precedes the kinematic inception threshold for the representative crests shown in Fig. \ref{fig:phase-diagram}. \citetalias{boettger2023} found that while both are generic features of the energetic evolution of a breaking wave, the energetic inception point precedes the kinematic inception point by around $0.1-0.2$ wave periods for the deep-water cases examined. We investigate the impact of wind on the timing of these breaking inception thresholds in Fig. \ref{fig:inception-times}. For each wind speed examined, the energetic ($t_e$) and kinematic ($t_k$) inception times are represented by their quartiles (box) and the minimum/maximum values within 1.5 times the interquartile range (whiskers). A representative $E_k$ evolution (dashed lines) is computed by regression to the median $E_k$ values at $t^* = [-1, t_e, t_k, 0]$. As the magnitude of wind forcing increases, the timing of the energetic inception point occurs earlier in the crest evolution, increasing from a median value of $t_e=-0.26t^*$ for $U_0 / c_p = 0$ to $t_e=-0.31t^*$ for $U_0 / c_p = 6$. A similar, although less distinct trend is also seen for the kinematic inception point with inception generally occurring earlier in wind-forced cases.
\section{Conclusions}

In this study, we have developed a numerical wave tank using a two-phase, volume-of-fluid Navier-Stokes solver to generate an ensemble of non-breaking and breaking waves under the influence of wind forcing. With this ensemble, which encompasses wind speeds ranging from moderate to strong wind forcing, we have examined the evolution of the terms in the kinetic energy balance equation (\ref{eq:DbEkDt-balance}) at the crest tip as the wave proceed towards breaking onset.  

The growth rate $D_b E_k / Dt$ of the kinetic energy at this point is dominated by the convergence of kinetic energy \con{} and its conversion to potential energy \KP{}. In non-breaking waves, these two terms are of approximately equal magnitude such that $D_b E_k / Dt$ remains small. Conversely, the evolution of breaking waves is characterised by a rapid increase in \con{} that is not offset by a corresponding increase in \KP{}, which results in a rapid increase in $D_b E_k / Dt$ up until breaking onset. The critical point in the evolution of the \con{} term which precedes the rapid increase in $D_b E_k / Dt$ was previously reported by \citetalias{boettger2023} and termed the energetic inception threshold. We found that this critical point was a generic feature of all breaking waves we examined in our ensemble. 

\citetalias{boettger2023} found that a threshold value of $D_b E_k / Dt = [0.198,0.235]$ exists, which must be exceeded for the deflection in the evolution of \con{} to result in breaking inception. We found this threshold value to be robust for our wind-forced wave experiments, despite the local kinetic energy at the crest tip $E_k$ increasing by up to 35 \% relative to an unforced wave of equivalent steepness. We also observed a correlation between the magnitude of \con{} and $D_b E_k / Dt$ and the breaking strength parameter $\Gamma$ \cite{derakhti2018, derakhti2020}, indicating that the energetic inception threshold can also quantify the strength of the subsequent breaking event. While wind forcing did not alter the threshold value, it did result in a systematic increase in the time between breaking inception and onset as the magnitude of the wind forcing was increased. 

The energetic breaking inception threshold provides new insights into the dynamical processes leading to wave breaking and offers opportunities for the parameterisation of the breaking impacts on larger scale ocean processes. The threshold has previously been shown to be robust for varying wave packet size in deep and intermediate water depths [\citetalias{boettger2023}], and our new results demonstrate that it is unaffected by wind forcing. Together, these results give confidence in the universality of the threshold, however, further studies examining a broader range of wave types are necessary to confirm these findings.  

\begin{acknowledgements}
This research was supported by the Commonwealth of Australia as represented by the Defence Science and Technology Group of the Department of Defence. Computational resources were provided through the National Computational Merit Allocation Scheme of the Australian Government's National Collaborative Research Infrastructure Strategy, as well as the University of New South Wales (UNSW) Resource Allocation Scheme. MLB acknowledges Australian Research Council support (grants DP120101701 and DP210102561) which motivated the present paper and its foundational contributions towards a more robust understanding of breaking water waves. 
\end{acknowledgements}

\bibliography{reference-library}

\end{document}